\documentstyle[twocolumn,aps]{revtex}
\tightenlines
\begin{document}

\newcommand{\beq}{\begin{equation}}
\newcommand{\eeq}{\end{equation}}
\newcommand{\bea}{\begin{eqnarray}}
\newcommand{\eea}{\end{eqnarray}}
\newcommand{\ba}{\begin{array}}
\newcommand{\ea}{\end{array}}
\newcommand{\om}{(\omega )}
\newcommand{\bef}{\begin{figure}}
\newcommand{\eef}{\end{figure}}
\newcommand{\no}{\nonumber}
\newcommand{\etal}{{\em et~al }}
\newcommand{\cf}{{\it cf.\/}\ }
\newcommand{\ie}{{\it i.e.\/}\ }
\newcommand{\eg}{{\it e.g.\/}\ }

\title{No-cloning theorem and teleportation criteria \\ 
for quantum continuous variables}
\author{Fr\'ed\'eric Grosshans and Philippe Grangier}
\address{Laboratoire Charles Fabry, Institut d'Optique Th\'eorique et Appliqu\'ee,
F-91403 Orsay, France}

\maketitle

\begin{abstract}

We discuss the criteria presently used for evaluating the efficiency of quantum teleportation
schemes for continuous variables. 
Using an argument based upon the difference between 1-to-2 quantum cloning (quantum duplication)
and 1-to-infinity cloning (classical measurement), we show that 
a fidelity value larger than 2/3 is required for successful 
quantum teleportation of coherent states. 
This value has not been reached experimentally so far.

\vspace{0.25cm}

\end{abstract}

\section {Introduction}

Quantum teleportation has emerged in recent years as a major paradigm of 
theoretical \cite{BBCJPW} and experimental \cite{BPMBBM,FLB98} quantum information. 
The initial approaches using discrete variables \cite{BBCJPW,BPMBBM} have been extended
to continuous quantum variables \cite{FLB98,BK98,RL98,dd}. 
However, various discussions have appeared recently about
the significance and the evaluation
criteria  of real, and thus imperfect, teleportation experiments \cite{BK98,RL98,dd,BK98b,qph1,qph2,err} . 

In this article, we will reconsider
the teleportation criteria for continuous quantum variables, 
with emphasis on the teleportation of coherent states \cite{FLB98}. 
Following the approach introduced in \cite{qph1,qph2}, we will show 
again that a fidelity $F_{tel} > 2/3$
is actually required for successful teleportation. 
In the present paper, our argument will be mostly based upon the no-cloning theorem 
for coherent states \cite{CIR,CI}.
We will show also how the present approach is related to the non-separability argument
that was used in \cite{qph1}. 

In order to set the scene, 
it may be useful to come back to \cite{BBCJPW}, 
where Bennett \etal introduce and define the concept of quantum teleportation.
This quotation is taken from their paper : 
`` Below, we show how Alice can divide the full information encoded in
[the unknown quantum state] $|\phi\rangle$ into two parts, one purely classical
and the other purely non-classical, and send them to Bob through two different
channels. Having received these two transmissions, Bob can construct an accurate
replica of $|\phi\rangle$.
{\em Of course Alice's original  $|\phi\rangle$ is destroyed in the process,
as it must be to obey the no-cloning theorem.}
We call the process we are about to describe teleportation, a term of science-fiction
meaning to make a person or object {\em disappear} while an exact replica appears
somewhere else."

From this definition, it should be clear that
teleportation has not only to beat the classical limits on measurement and transmission,
but also to enforce the no-cloning limit, otherwise Bob may receive a state
which is better than any classical copy, but nevertheless it will not be the teleported $|\phi\rangle$.
A crucial point is then that there is a distinction between non-clonable quantum information
and classical information. This is best illustrated by considering the 
fidelity for cloning one copy of a coherent state into $M$ identical copies,
which is $F_{1\to M} = 2/(2+N^{ad})$, where $N^{ad} = 2 (M-1)/M$, as shown by Cerf and Ilbrisdir \cite{CI}
($N^{ad}$ is an equivalent noise in the cloning process \cite{qph1}, that will be discussed in more detail below).
It is then clear that $F_{1\to\infty} = 1/2$, while $F_{1\to 2} = 2/3$. 
The usual criterion about teleportation assumes correctly that a classical measurement
is involved in teleportation. However, it concludes incorrectly that the relevant limit is thus
the one associated with a classical measurement, $F_{1\to\infty} = 1/2$. 
This conclusion is incorrect because the good question to ask is :
what is the measured fidelity of Bob's copy, as measured by the verifier Victor,
which warrants that no better copy of the input state can exist elsewhere ?
(\ie, kept by a cheating Alice, or eavesdropped by a malicious Eve). 
We will show below in detail, but it should already be
clear from the above cloning limit, that the correct answer is $F_{1\to 2} = 2/3$. 

\section{The $1\to 2$ and $1\to M$ cloning limits}

\subsection{Quantum duplication}
\label{Sec12Cloner}

We first give a simple demonstration that the fidelity limit for making two copies
of an input state is $F_{1\to 2} = 2/3$, as it was previously shown by Cerf \etal in \cite{CIR}.
Here we recover the same conclusion by using simple techniques similar to the ones
used for evaluating QND measurements, introduced in \cite{HCW90,PRG94,GLP} and used in \cite{RL98,qph1,qph2}.

A $1\to 2$ cloner or `duplicator' has one input mode and two output modes $a$ and $b$.
Denoting by $g$ and $B$ the (linearized) gains and noises for each channel,
the quadratures of the two output modes are related to the two input quadratures $X_{in}$
and $Y_{in}$ by

\begin{eqnarray}
    X_a&=&g_{X_a}X_{in}+B_{X_a} \; \; \; \; Y_a=g_{Y_a}Y_{in}+B_{Y_a} \nonumber \\
    X_b&=&g_{X_b}X_{in}+B_{X_b} \; \; \; \; Y_b=g_{Y_b}Y_{in}+B_{Y_b}
\label{xyab}
\end{eqnarray}

Since $a$ and $b$ are two different field modes, any observable of $a$
commutes with any observable of $b$, and in particular  $[X_a,Y_b]=0$.
Using equations (\ref{xyab}), and assuming that 
the added noises are not correlated to the input signals, we obtain :
\begin{equation}
[B_{X_a},B_{Y_b}]=-g_{X_a}g_{Y_b}[X_{in},Y_{in}],
\end{equation}
The noises added by the duplicator verify therefore
\begin{equation}
\Delta B_{X_a}\Delta B_{Y_b}\ge|g_{X_a}g_{Y_b}| N_0,
\end{equation}
where $N_0$ is the vacuum's noise variance
and $\Delta$ denotes the usual rms dispersion.

It is convenient to define the variances of the 
equivalent input noises \cite{PRG94} associated with the measurements :
\bea
N_{X_{i}} = (\Delta X_{i}/|g_{X_{i}}|)^2  - (\Delta X_{in})^2 = (\Delta B_{X_{i}}/|g_{X_{i}}|)^2 \nonumber \\
N_{Y_{i}} = (\Delta Y_{i}/|g_{Y_{i}}|)^2  - (\Delta Y_{in})^2 = (\Delta B_{X_{i}}/|g_{X_{i}}|)^2 
\eea
where $i$ is either $a$ or $b$. One obtains thus the symmetrical inequalities : 
\beq
N_{X_a} N_{Y_b} \geq N_0^2 \; \; \; \; \; \; 
N_{X_b} N_{Y_a} \geq N_0^2
\label{EqHICloning12}
\end{equation}
These inequalities are very similar to the ones that appear in QND measurements
\cite{PRG94}, and they ensure that building two copies of the input state
will not allow one to work around the Heisenberg inequality. Actually, the added noise 
is just the one required to forbid to infer the values of $X_{in}$ and $Y_{in}$ with
a precision better than Heisenberg limit, by measuring $X_a$ and $Y_b$.

The equivalent noises can be easily related to the cloning fidelity.
Actually, it was shown in \cite{qph1,qph2} that 
the fidelity  obtained when copying coherent states with unity gain ($g_{X_i}= g_{Y_i} =1$) is
given by :
\beq
F_{g_T=1} = \frac{2}{\sqrt{(2+N_X/N_0)(2+N_Y/N_0)}}
\label{deff}
\eeq
Assuming that the two copies are identical and have phase-independant noise, 
the limit of equation (\ref{EqHICloning12}) is reached for 
$N_{X_a}=N_{Y_b} =N_{X_b}= N_{Y_a}= N_0$ and corresponds thus 
to $F_{g_T=1} =2/3$. This is identical to the result obtained by Cerf \etal in \cite{CIR}.
A `duplicator' reaching the limit of equation
(\ref{EqHICloning12}), can be easily implemented using a linear amplifier and a 50/50 beamsplitter.
Such a duplicator is a gaussian cloning machine as defined by Cerf \etal \cite{CIR}.
Various implementations of ``cloners" have been proposed recently  \cite{cloners},
and may allow in particular to share arbitrarily the
noise between one copy with is kept, and another one which is sent out.

\subsection{The $1\to M$ cloning limit}

We generalize here the above demonstration to copying one input to $M$ identical outputs.
In order to recover directly the result of Cerf and Iblisdir \cite{CI},
we will assume that each output channel has unity gain, and that all copies are identical 
in the sense that the variances are the same for all output,
and that the pairwise correlation does not depend on the pair of outputs which is considered.
More precisely, the quadratures of the $M$ outputs of a $1\to M$ cloner ($M>2$)
obey
\begin{equation}
\left\{
                \begin{array}{l}
    X_i=X_{in}+B_{X_i}\\
    Y_i=Y_{in}+B_{Y_i},
                \end{array}
\right.
\end{equation}
for every $1\le i\le M$.
We define $C_X$, $N_X$ and $N_Y$ as :
\begin{eqnarray}
C_X&=&\langle B_{X_i} B_{X_j}\rangle \hbox{ for every } i\neq j \no \\
N_X&=&\Delta B_{X_i}^2 \mbox{ for every } i \no \\
N_Y&=&\Delta B_{Y_i}^2 \mbox{ for every } i.
\label{eqM}
\end{eqnarray}
Like in section \ref{Sec12Cloner}, we have
\begin{eqnarray}
\label{EqCommutDiff}
[B_{X_i}, B_{Y_j}]&=&-[X_{in},Y_{in}] \hbox{ for every } i\neq j\\
\label{EqCommutId}
[B_{X_i},B_{Y_i}]&=&0 \mbox{ for every } i
\end{eqnarray}
We can define $\Lambda$ for any real number $\lambda$ by
\begin{equation}
\label{EqDefLambda}
\Lambda= B_{X_1} + \lambda\sum_{i=2}^MB_{X_i}.
\end{equation}
It follows straightforwardly from equations (\ref{EqCommutDiff}) and (\ref{EqCommutId}),
that
\begin{equation}
[\Lambda,B_{Y_1}]=-\lambda(M-1)[X_{in},Y_{in}].
\end{equation}
For the variances, it implies
\begin{equation}
\label{IneqLambda}
\Delta\Lambda\Delta B_{Y_1}\ge|\lambda(M-1) N_0|.
\end{equation}
Computing $\Delta\Lambda^2$ directly from eq. (\ref{EqDefLambda}), we have
\bea
\label{EqVarLambda}
\Delta\Lambda^2 &=& \Delta B_{X_1}^2 + \lambda^2\sum_{i=2}^M\Delta B_{X_i}^2 \no \\
                 &+& 2\lambda\sum_{i=2}^M\langle B_{X_1} B_{X_i}\rangle 
 +\lambda^2\sum_{i,j>1}^{i\neq j}\langle B_{X_i} B_{X_j}\rangle.
\eea
Using the definitions (\ref{eqM}) in eq. (\ref{EqVarLambda}), we obtain
\bea
\Delta\Lambda^2 &=& [1+\lambda^2(M-1)] N_X \no \\
                 &+& [2\lambda(M-1)+\lambda^2(M-1)(M-2)]C_X.
\eea

If $\lambda=-2/(M-2)$, this expression is simpler and becomes
\begin{equation}
\Delta\Lambda^2=\frac{M^2}{(M-2)^2} N_X,
\end{equation}
which can be injected in eq. (\ref{IneqLambda})
to obtain the $1\to M$ cloning limit
\begin{equation}
\label{HILimit1M}
N_X N_Y\ge \left( \frac{2(M-1)}{M} \right)^2 N_0^2
\end{equation}
This limit is also valid for $M=2$ as written in eq.(\ref{EqHICloning12})
and for the trivial case $M=1$.

Assuming that the $M$ copies have phase-independant noise, 
\ie $N_X/N_0=N_Y/N_0=N^{ad}=2  (M-1)/M$, it is simple to show from eq. \ref{deff} 
that the corresponding fidelity limit for coherent state cloning is
\begin{equation}
F_{1 \to M} =\frac{2}{2+N^{ad}} \le\frac{M}{2M-1}.
\end{equation}

As previously, a perfect $1\to M$ symmetrical cloner can be implemented
using a linear amplifier and $M-1$ beamsplitters \cite{cloners}.

\subsection{The $1\to \infty$ cloning and classical measurements}

When a classical measurement is performed, the measurement result can be copied
an arbitrary number of time. It should thus be clear that the limit corresponding to a classical
measurement is $F_{1 \to \infty} = 1/2$, or $N^{ad}=2$. On the other hand,
making only two copies comes at a smaller price, and corresponds to 
 $F_{1 \to 2} = 2/3$, or $N^{ad}=1$. We will show below that this distinction is crucial as far as 
quantum teleportation is concerned.

\section{Teleportation and no-cloning}

\subsection{Quantum teleportation criteria}

Suppose Alice (a) sends a quantum state to Bob (b),
who wants to be certain that Alice cannot have kept a better copy
of the input state than the one she has given to him. 
This requirement means to be sure that Alice's copy is destroyed,
\ie that real quantum teleportation has occurred. 
Alice will be able to cheat if her equivalent noise is smaller than
Bob's, that is :
\begin{equation}
N_{X_b} \geq N_{X_a}^{opt} \mbox{ and } N_{Y_b} \geq N_{Y_a}^{opt}.
\end{equation}
where $opt$ denotes the optimum result for Alice. 
Since the best Alice can do is limited by the Heisenberg-like inequalities (\ref{EqHICloning12}), one has :
\begin{equation}
N_{X_b} \geq N_0^2 / N_{Y_b}\mbox{ and } N_{Y_b} \geq N_0^2 / N_{X_b}.
\end{equation}
and thus 
\begin{equation}
N_{X_b}N_{Y_b} \geq N_0^2.
\end{equation}
If Bob's noise variances are  symmetrical, \ie $N_{X_b}=N_{Y_b}$,
one recovers the limit 
\begin{equation}
F \le 2/3
\end{equation}
for teleporting coherent states.
Thus the only way for Victor to warrant that Alice is not cheating is 
to obtain a measured teleportation fidelity larger that $2/3$.
It is worth noticing that when the associated condition $N_{X_b} N_{Y_b} < N_0^2$ 
is fulfilled, then eq. (\ref{EqHICloning12}) imposes that 
$N_{X_b} < N_{X_a}$  and $N_{Y_b} < N_{Y_a}$, and thus Alice will have both quadratures
worse than Bob.

\subsection{Security in quantum teleportation}

It should be clear now that as long as $F\le 2/3$, Alice can cheat teleportation 
by keeping a better copy than the one Bob has received. The simplest way to do that is first
to duplicate the input state, then to keep
one copy, and to teleport the other one to Bob. As an example, if Bob's teleported
output has a fidelity $F_b = 0.58$, or $N_b = 1.45$, and if Alice has a perfect teleporter than
she claims to be imperfect, she can keep a copy with a fidelity $F_a = 0.74$, or $N_a = 0.7$. 
This is clearly not acceptable according to the definition of \cite{BBCJPW}. 

We point out that the same condition applies when Alice is honest, but 
when quantum teleportation is used to send a quantum state from Alice to 
Bob for quantum cryptography purposes. In that case, one must worry about the
amount of information which can be eavesdropped during the teleportation process. 
For simplicity, let us consider a teleportation scheme using EPR beams, 
with a finite degree of squeezing, 
and transmission losses. It is assumed that Eve is able to perfectly eavesdrop the classical channel, 
and that she has full access to the losses along at least one ``transmission arm" of the EPR beam 
(this is  a strong hypothesis, but it is usually done for evaluating the security of standard 
quantum cryptography). 
The simplest solution for Eve is to build her own teleported state, and she will be
successful if this state has an equivalent noise 
smaller than the one achieved by Bob. 
It can be shown simply, and it is physically obvious, that as long as
the EPR channel efficiency $\eta$ is smaller than $1/2$, 
Eve can obtain a teleported copy of the input state which is {\it better}
than the one obtained by Bob. 
More generally, this can be also seen as a consequence of the $1 \to 2$ cloning limit :
if $F$ is larger than $2/3$, Bob can be sure
that a malicious Eve will not be able to eavesdrop the teleported state \cite{qph2}. 
Thus the $F>2/3$ limit appears also as a crucial
security condition if teleportation is used
as a quantum communication tool.

\subsection{Discussion}

In order to clarify the issues involved, it may be worth summarizing the physics
involved in the respective criteria $F>1/2$ and $F>2/3$.

As said above, $F=1/2$ is actually a classical measurement limit,
directly associated with the $1 \to \infty$ cloning limit.
It has also been shown in  \cite{pp2} that purifications procedure
can be initiated as soon as $F>1/2$, and may lead to high fidelity values.
However, the purpose of teleportation criteria is to characterize 
a given experiment, and not what it might be by adding purification procedures. 
We note also that recently demonstrated entanglement criteria \cite{PH}
are fully compatible with the $F=1/2$ limit. 
It is thus clear that the $F>1/2$ criterion characterizes a threshold for 
the appearance of non-classical effects in the teleportation process \cite{err}. 
However, this does not seem to be
enough to warrant that successful teleportation has been achieved.

On the other hand, 
the main virtue of the  $F>2/3$ criterion is that it warrants that no other copy
of the input state
can remain, that would have a better fidelity than the one Bob has received.
This results from the no-cloning theorem, which is in turn related to the ability
to infer one result on one system, given a measurement done on another system.
This point is closely related to the EPR
non-separability argument of ref. \cite{qph1}, which 
requires that ``conditional squeezing" can be obtained on one EPR beam,
given a measurement that is done on the other one.
In particuler, in a EPR teleportation scheme,  the intensity of one EPR 
beam can be measured in order to use that information to reduce the noise
of the second beam  \cite{ens}. 
Then the noise of the corrected beam can be reduced below shot-noise
only when the losses on each beam are less than 50\% \cite{qph1,qph2}. 
A central point, which was shown in \cite{qph1}, is that  $F>2/3$ is a sufficient condition
to warrant that such ``conditional squeezing" can be obtained.
The $F=2/3$ limit is thus closely related
to inferences made using conditional variances, which also play an essential role
in QND measurements \cite{HCW90,PRG94,GLP}.

Finally, we propose to call the uncertain region between $F=1/2$ and $F=2/3$
the ``quantum fax" region (see Figure \ref{fig1}). 
This means that a quantum entanglement resource
must be used to reach that region, 
but nevertheless that full quantum teleportation has not been completed,
because the no-cloning theorem is not yet enforced. Therefore, as in a fax machine,
Bob has received something which is not so bad, but a better copy may still exist somewhere.
Obviously, the region above $F=2/3$ is full quantum teleportation.

\section{Conclusion}

As a conclusion, it should be clear that the criteria $F>1/2$ and $F>2/3$
have different physical contents, and are both legitimate.
Based upon the definition given in \cite{BBCJPW}, and on the no-cloning
theorem, we claim that full teleportation requires $F>2/3$. 
However, it should be clear that though the result $F_{exp} = 0.58$ reported
in ref. \cite{FLB98} falls below that value, this experiment is nevertheless a very significant
achievement in defining and using the concept of continuous variables quantum teleportation. 

\section*{Acknowledgements}

This work was carried out in the framework of the european IST/FET/QIPC project ``QUICOV".


\begin{figure}
\vspace{4.75cm}
\vspace{0.5cm}
\caption{Table illustrating the fidelity values
associated to the number of allowed copies of the input state.
The value $F=1/2$ corresponds to the threshold for quantum
effects, while the value $F=2/3$ corresponds to the enforcement of no-cloning. 
For distant operations, the region between $F=1/2$ and $F=2/3$ is called
``quantum fax", because an original may be kept by Alice, with a higher fidelity
than the teleported copy. The no-cloning region $F>2/3$ corresponds to tranfering
quantum states between different systems for local operations, and
to quantum teleportation for remote operations. }
\label{fig1}
\end{figure}


\begin{references}


\bibitem{BBCJPW}  C.H.~Bennett \etal, Phys. Rev. Lett. {\bf 70}, 1895 (1993).

\bibitem{BPMBBM}  
D. Bouwmeester \etal, Nature {\bf 390}, 575 (1997);
D. Boschi \etal, Phys. Rev. Lett. {\bf 80}, 1121 (1998).

\bibitem{FLB98}
A. Furusawa \etal, Science {\bf 282}, 706 (1998).

\bibitem{BK98}
S.L. Braunstein and H.J. Kimble, Phys. Rev. Lett. {\bf 80}, 869 (1998); 
S.L. Braunstein, C.A. Fuchs and H.J. Kimble, J. Mod. Opt. {\bf 47}, 267 (2000).

\bibitem{RL98} 
T.~C.~Ralph and P.~K.~Lam, Phys. Rev. Lett.  {\bf 81}, 5668 (1998);
T.C.~Ralph, Opt. Lett. {\bf 24}, 348 (1999).

\bibitem{dd}
L. Vaidman and H. Yoran, Phys. Rev. A {\bf 59}, 116 (1999);
S.M. Tan,  Phys. Rev. A {\bf 60}, 2752 (1999);
S.J. van Enk, Phys. Rev. A {\bf 60}, 5095 (1999);
P. van Loock  and S.L. Braunstein, Phys. Rev. A {\bf 61} 010302 (2000).

\bibitem{BK98b}
S.L. Braunstein and H.J. Kimble, Nature {\bf 394}, 840 (1998).

\bibitem{qph1} P. Grangier and F. Grosshans, 
``Quantum teleportation criteria for continuous variables", arXiv:quant -ph/0009079.


\bibitem{qph2} P. Grangier and F. Grosshans, 
``Evaluating quantum teleportation of coherent states", arXiv:quant-ph/0010107. 

\bibitem{err} S.L. Braunstein, C.A. Fuchs, H.J. Kimble, P. van Loock,
``Quantum versus Classical Domains for Teleportation with Continuous Variables",
arXiv:quant-ph/0012001.

\bibitem{CIR}  N.J.~Cerf, A.~Ipe and X.~Rottenberg,
            Phys. Rev. Lett. {\bf 85}, 1754 (2000).

\bibitem{CI}  N.J.~Cerf and S.~Iblisdir,
            Phys. Rev. A {\bf 62}, 040301 (2000).

\bibitem{HCW90} M.~J.~Holland, M.~J.~Collett, D.~F.~Walls and
M.~D.~Levenson, Phys. Rev. A {\bf 42}, 2995 (1990).

\bibitem{PRG94} J.~-Ph.~Poizat, J.~-F.~Roch and P.~Grangier,
Ann. Phys. Fr. {\bf 19}, 265 (1994).

\bibitem{GLP} Ph. Grangier, J.-A. Levenson and J.-Ph. Poizat, 
	Nature {\bf 396}, 537 (1998).

\bibitem{cloners} For several recent proposals of cloning devices see \eg
arXiv:quant-ph/0012025; ibid/0012046; ibid/0012048


\bibitem{pp1} S. Parker \etal, Phys. Rev. A {\bf 61}, 032305 (2000)

\bibitem{pp2} L.-M. Duan \etal, Phys. Rev. Lett. {\bf 84}, 4002 (2000).

\bibitem{PH} L.-M. Duan \etal, Phys. Rev. Lett.  {\bf 84}, 2722 (2000);
R. Simon, Phys. Rev. Lett.  {\bf 84}, 2726 (2000).

\bibitem{ens} J. Mertz \etal, Phys. Rev. A {\bf 44}, 3229 (1991)

\end{references}
\end{document}